\documentclass{iaus}
\usepackage{graphicx}

\title[The Thin Disk] 
{The Stellar Population of the Thin Disk}

\author[Carlos Allende Prieto]   
{Carlos Allende Prieto$^1$
\thanks{Present address: Instituto de Astrof\'{\i}sica de Canarias, V\'{\i}a L\'actea S/N, 
La Laguna 38205, Tenerife, Spain.},
}

\affiliation{$^1$Mullard Space Science Laboratory, University College London, \\ Holmbury St. Mary,
RH5 6NT , Surrey, United Kingdom \\ email: {\tt callende@astro.as.utexas.edu}}

\pubyear{2009}
\volume{265}  
\pagerange{119--126}
\jname{Chemical Abundances in the Universe: Connecting First Stars to Planets}
\editors{K. Cunha, M. Spite \& B. Barbuy, eds.}
\begin{document}

\maketitle

\begin{abstract}

We discuss recent observations of stars located close to
the symmetry plane of the Milky Way, and examine them in the context of
theories of Galaxy formation and evolution. The kinematics, ages, and compositions
of thin disk stars in the solar neighborhood display complex patterns, and
interesting correlations. The Galactic disk does not seem to pose 
any unsurmountable obstacles to hierarchical galaxy formation theories, but 
a model of the Milky Way able to reproduce the complexity found in the data 
will likely require a meticulous study of a significant fraction of the stars
in the Galaxy. Making such an observational effort seems necessary in order 
to make  a physics laboratory out of our own galaxy, and ultimately 
ensure that the most relevant processes are properly understood.

\keywords{Keyword1, keyword2, keyword3, etc.}
\end{abstract}

\firstsection 

\section{Introduction}

Large galaxies naturally produce disks. Radiative cooling of the gas and
angular momentum conservation lead the early evolution of galaxies through 
dissipative collapse and disk formation. Disks are frequently observed 
in galaxies,  even at high redshift (F\"orster Schreiber et al. 2006), and
the Milky Way does not seem unique at all in showing a {\it dual} disk, with
a distinct thin and thick components (Dalcanton \& Bernstein 2002).

Early attempts to place the Galactic thin disk in the context of 
a $\Lambda-$CDM universe exposed a number of problems. The number
of observed surviving satellites appeared far too small compared to
simulations. This problem is somewhat alleviated after the Sloan Digital Sky Survey 
(SDSS) has identified many new low-surface brightness galaxies 
in the immediate Galactic neighborhood (but see Koposov et al. 2008). 
It was also deemed hard for the disk to survive for as long as 
the observations suggested,  8--10 Gyr, and in particular to stay thin 
(T\'oth \& Ostriker 1992, Kauffmann \& White 1993).
More recent appraisals, however, indicate that as many as 85\%
of disk galaxies have not been involved in a merger with
a mass ratio larger than 0.5 since redshift $\sim 1-1.5$, or approximately 
in the last 8 Gyr
(Koda et al. 2009).
A higher gas fraction in the  accreted building blocks seems to 
favor disk survival (Hopkins et al. 2009).

The structure of galaxy disks is usually studied measuring
surface brightness distributions. Anomalies such as spiral arms
and HII regions are smoothed out, taking de-projected azimuthal averages
in nearly face-on galaxies, modeling the radial dependence of the light 
distribution with exponential profiles
(see, e.g., Aguerri et al. 2000; Prieto et al. 2001). Edge-on galaxies are, 
in turn, used to study the light distribution perpendicular to the plane.
In the Milky Way, the spatial distribution of stars is studied using deep 
imaging surveys -- counting
stars and exploiting photometric calibrations to estimate 
the luminosity of the main-sequence as a function of color 
(e.g. Juri\'c et al. 2008). 

Being {\it insiders} to the Galaxy provides some advantages; 
for example, we can measure in detail the
properties of individual stars using spectroscopy.
Modern surveys employ cameras with a very broad 
dynamical range, and massive multiplexing capabilities 
for spectroscopy (see, e.g., Gunn et al. 1998, 2005; Onaka et al. 2008),
 making it feasible to obtain large data sets fast.

\section{Main structural components of the Milky Way}

There are multitude of studies of the 
main Galactic components using star counts.
Two recent studies by Cabrera-Lavers et al. (2007) and Juri\'c et al.
(2008) based on 2MASS (Skrutskie et al. 2006) and the SDSS 
(Abazajian et al. 2009), respectively, sample quite well the
Galactic thin disk. These surveys are
dominated by late-type (mainly K and M dwarf) stars, and
their scale height is found to be about 200--300 pc. There
is no consensus on the (radial) scale length of
the thin disk, and estimates range between 2.5 to 3.5 kpc, but
the larger distances involved make this 
measurement harder. The thin disk is thought to contribute about 
85\% of the stars in the Galactic plane.

It is important to emphasize that the disk scale heights are
expected to vary depending on where we look 
(Bilir et al. 2008), as the potential is far from perfectly
smooth and axisymmetric. More dramatic is the variation of the scale
height of the thin disk with age. Thus, Ma\'iz-Apellaniz (2001)
finds $h\sim 35$ pc from OB type stars -- a value nearly 10 times smaller
than determinations from late-type stars. This strong age dependence is also
imprinted in the distribution of M-dwarfs observed spectroscopically
in the SDSS, and West et al. (2004, 2006, 2008) find  a decreasing
fraction of active stars (the youngest) as they sample farther away from the 
Galactic plane. Further evidence is also seen in the stellar kinematics
in the solar vicinity, which are discussed in the next section.

\section{Stellar kinematics in the Solar Neighborhood}

The combination of Hipparcos (and Tycho) astrometry with radial velocities
from high-resolution spectroscopy has provided 
detailed coordinates in phase space for stars in the solar neighborhood 
($< 100$ pc from us).  Observations with the CORAVEL spectrographs 
(Baranne et al. 1979) have been used to study large samples of 
giants (Famaey et al. 2005) and especially F- and G-type dwarfs
(the Geneva-Copenhagen survey, described in 
Nordstr\"om et al. 2004 and Holmberg et al. 2007, 2009).

The velocity distributions of nearby stars show plenty of
structure. We discuss structure in more detail in the following
section, and here we focus on other characteristics. 
Each of the velocity components of the thin disk ($U$, $V$ and $W$ 
for the radial, azimuthal, and vertical components in a cylindrical
coordinate system; see Fig. \ref{fe}) shows distributions that increase in width 
with age. The scatter in the vertical velocities ($W$) shows
the sharpest and smoothest rise with age of the three velocity components
from $\sigma \sim 8$ km s$^{-1}$
for stars that are $\sim 1$ Gyr old to roughly 30 km s$^{-1}$
for stars $\sim 10$ Gyr old. This is another way of looking at the 
increase in scale height with time.

For a Mestel disk (surface density $\Sigma \propto R^{-1}$, where
$R$ is the galactocentric distance),
assuming an isothermal sheet $\rho \propto \cosh^{-2}(z)$, embedded 
on a spherical halo 
($\rho \propto r^{-2}$, where $r$ is the radial coordinate), the
disk rotational velocity $V_{\rm rot}$ is constant, and the vertical velocity 
dispersion can be written
\[
\sigma^2 = \frac{1}{2} V_{\rm rot}^2 \frac{h}{R_{\odot}} 
\left( \mu + \lambda(1-\mu) \frac{h}{R_{\odot}} \right),
\]
\noindent (T\'oth \& Ostriker 1992),   
where $\lambda \equiv \int x^2 \cosh^{-2}(x) dx \simeq$ 1.645,
$\mu(R)$ is the enclosed mass ratio (disk/total, about 0.35 for the Milky Way), 
and adopting  
$V_{\rm rot}\simeq 220$ km s$^{-1}$, 
\[
\sigma^2 \simeq 24200 \frac{h}{R_{\odot}} \left(0.35 + 1.10 \frac{h}{R_{\odot}}\right).
\]
\noindent This indicates that a range in $\sigma$ 
between 8 and 30 km s$^{-1}$, as observed for stars between 1 and 10 Gyr
should be matched by an excursion in the scale height between 60 and
600 pc. Measuring directly scale heights
from astrometry will have to wait for Gaia (Lindegren \& de Bruijne 2005; 
Lindegren et al. 2009), as the Hipparcos
parallaxes for late-type dwarfs are limited to $\sim 100$ pc.

\section{Structure in the Galactic disk}

Eggen is often singled out as the pioneer of the study of comoving 
groups of stars (superclusters and moving groups) 
in the solar vicinity  (see, e.g., Eggen 1992). Recent
data sets have shown these structures with sharper contrast, and revealed
new ones (see, e.g. Famaey et al. 2005; Arifyanto \& Fuchs 2006). Large 
surveys such as RAVE (e.g. Klement et al. 2008), and ultimately Gaia, 
are to provide improved statistics that will 
bring light on the important topic of the origin of 
superclusters and their connection to field stars and proper clusters.
 
Previous work has demonstrated that some superclusters are indeed dissolving
stellar clusters (e.g., the HR 1614 moving group, which has
been found to exhibit a single age and metallicity by De Silva et al. 2007).
But many others that have been scrutinized appear to exhibit broad
age spans (e.g. the Pleiades, Hyades, or Hercules superclusters;
Famaey et al 2008), or chemical abundance distributions (e.g. Hercules;
Bensby et al. 2007), which suggests they have a dynamical origin (e.g. De Simone
et al. 2004; Quillen \& Minchev 2005; Chakrabarty 2007). Note that the 
accretion of an external stellar system may offer, in some cases, a
plausible formation scenario, with stars directly brought in with common 
kinematics, or simply 'linked' as a result of an accretion
event (Minchev et al. 2009; Quillen et al. 2009).

\section{Ages and metallicities of disk stars}

Being able to date individual stars is extremely valuable. 
The recovery of the
star formation history of the stars in the solar neighborhood
from the inversion of observed HR diagrams or chromospheric age 
estimates has been attempted
in many studies (Rocha-Pinto et al. 2000; Hern\'andez et al. 2000; 
Bertelli \& Nasi 2001; Aumer \& Binney 2009).  Unfortunately, an examination  
of these and other works does not provide a coherent picture.

Isochrone dating is mostly limited to subgiants, as it is 
at the turn-off where the basic fundamental parameters, mainly 
the luminosity, change quickly, with minimal degeneracy; i.e.
isochrones spread nicely. Surface gravity determinations from
spectroscopy are hardly useful, since spectra are only weakly
sensitive to pressure, and fundamental measurements such as 
trigonometric parallaxes and angular diameters are best to
constrain stellar luminosities. Gaia will dramatically change
this field with parallaxes accurate to $\sim 20$ $\mu$m at 15 magnitude.

Paying attention to details, in particular applying a rigorous
statistical analysis, is important, and in some extreme cases
critical. The last few years have seen a change   
in the methodologies for determining stellar ages,
from the crude method of assigning the nearest isochrone to  
sophisticated
statistical analyses (see, e.g., Reddy et al. 2003; Pont \& Eyer 2004;
J{\o}rgensen \& Lindegren 2005). 
It has been emphasized by Lachaume et al. (1999) that different
dating techniques are complementary. For example, isochrones
are most useful for turn-off stars, and hence more likely applicable
to intermediate mass stars, while activity and rotation 
can provide ages for low mass stars that stay on the main sequence 
 longer than a Hubble time.

Thin disk stars show a wide range of ages. Reddy et al. (2006)
estimated ages between 1 and 9 Gyr, although predominantly $<5$ Gyr. 
Holmberg et al. (2009) and Haywood (2008), using larger samples,  found   
wider ranges reaching 
up to 13--14 Gyr, although most concentrated again at $<4-5$ Gyr.
Very old ages for thin-disk stars may be at odd with the upper limit to
the age of the disk derived from the analysis 
of the white dwarf cooling sequence. For example, Leggett al. (1998)
estimated $8 \pm 1.5$ Gyr, although a critical
assessment of the literature by Fontaine, Brassard \& Bergeron (2001)
led them to propose a plausible range between 8.5 to 11 Gyr.

\begin{figure}[t]
\begin{center}
 \includegraphics[width=3.4in,angle=90]{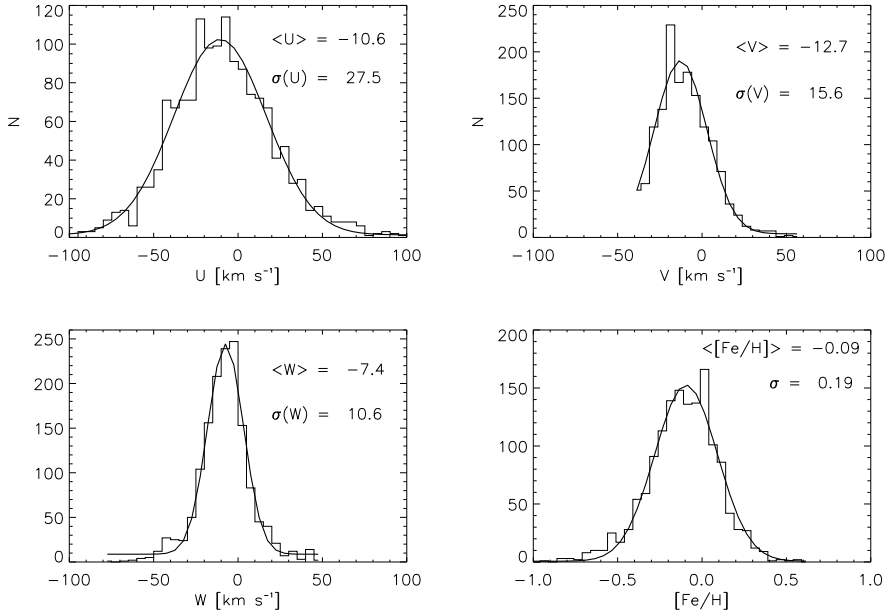} 
 \caption{Velocity and metallicity histograms for 2427 stars  with metallicities
 from the catalog of Cayrel de Strobel, Soubiran \& Ralite (2001), Hipparcos
 astrometry, and radial velocities from the compilations by 
 Malaroda, Levato \& Galliani (2001) and/or 
 Barbier-Brossat \& Figon (2000),
 showing  velocities $V> -40$ and $-80 < W < +50$ km s$^{-1}$. 
 The expected contamination by thick disk and halo populations is very small, 
 and therefore we identify the observed distributions with the thin disk. The
 smooth solid lines are Gaussian curves fitted to the histograms. Note the peak
 associated with the Hyades at ($U,V,W$) $\simeq$ ($-43, -18, -2$) km s$^{-1}$.
 Adopted from Allende Prieto et al. (2004).}
   \label{fe}
\end{center}
\end{figure}

An inspection of the literature in the last decade shows
that there is now consensus regarding the 
metallicity distribution of the thin disk. 
Most authors find it is reasonably
Gaussian,  with a standard
deviation of about 0.2 dex (see, e.g. Allende Prieto et al. 2004;
Holmberg et al. 2007). More polemic is the exact 
mean of the distribution, which 
some argue could be as low as [Fe/H]=$-0.10$ dex, while others push for
a value much close to solar (hence around $0.00$; see Luck \& Heiter 2007; 
Haywood 2001; Taylor \& Croxall 2005; Fuhrmann 2008).

As noted some years ago, it is interesting that blindly adopting
the metallicities compiled in the Cayrel de Strobel et al. catalog, 
and just by simply cleaning thick disk stars with {\it slow}
Galactic rotation ($V<$ 50 km s$^{-1}$\footnote{Thick disk stars lag behind the thin disk 
rotation by roughly that much, although this depends on the distance from the plane.}), 
one recovers again a
[Fe/H] distribution centered at $-0.1$ dex with $\sigma\simeq 0.20$ dex
-- remarkably close to those found from the analysis of much more
homogeneous data sets (see Fig. \ref{fe}).
The implication is that despite this compilation includes high-resolution
determinations from many studies, the systematics across samples
and analysis protocols are not large enough to widen the 
derived distribution.

An additional complication should be noted. Although the metallicity
distribution of the thin disk may be well determined, this is likely
not an ideal quantity to compare with chemical evolution models.
Selection effects such as mass biases due to different lifetimes
need to be considered, and so does 
the role of radial migration, which could be bringing significant
numbers of stars formed at different galactocentric distances, and
hence with different compositions even if formed at exactly the same time
(see, e.g., Haywood et al. 2008).

\section{Abundance ratios}

There have been a multitude of studies performing high-resolution 
spectroscopy
of nearby GFK stars (e.g., Edvardsson et al. 1993; Feltzing \& Gustafsson 1998; 
Chen et al. 2000;
Nissen et al. 2000; Fulbright 2002; Reddy et al. 2003, 2006; Takeda 2007; 
Ecuvillon et al.
2004; Gilli et al. 2006; Bensby et al. 2003; Ram\'{\i}rez et al. 2007; 
Fuhrmann 1998, 2004, 2008).  
Most of  these studies found a remarkable uniformity in 
the abundance ratios for thin disk stars at any given [Fe/H]. 

Reddy et al. (2003) looked for and failed to find a cosmic scatter in the
abundance ratios. Assuming
[Fe/H] is a reliable clock, 
the interstellar medium where these samples formed was very well mixed.
Many works encounter non-solar ratios at solar [Fe/H] for some elements.
This puzzling result, which might fuel the idea of the Sun being somehow 
special, was later traced to systematic
errors in the abundances associated with using the Sun as a reference for
non-solar type stars (Allende Prieto 2008). There is no doubt 
that highly homogeneous samples, in particular those restricted to
a narrow range in  effective temperature ({\it isothermal} samples, if you will),
can dramatically reduce systematic errors still present in the analyses.
Mel\'endez et al., in these proceedings, show an extreme example of exploiting 
such a trick.

\section{A dichotomy between the thin and thick disks?}

The thin and thick disks stars in the solar neighborhood 
can be easily separated, at
least statistically. Although the distributions of the velocity
components and metallicities overlap somewhat, combining all the data
for $UVW$ as well as [Fe/H], makes their separation fairly straightforward.
The age distributions have probably very little overlap, if any at all
(see, e.g., Fig. 24 in Reddy et al. 2006).
Star formation in the Milky Way has likely proceed in phases,
with limited overlap: halo, thick disk, and thin disk. Yet, 
the connection between these three components, and in particular
the thick and thin disks, is far from understood.

\begin{figure}[t]
\begin{center}
 \includegraphics[width=3.4in]{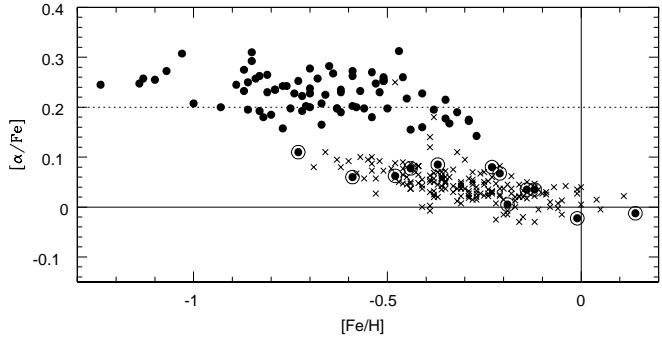} 
 \caption{Average of the abundance ratios of Mg, Si, Ca and Ti to Fe for 
 stars kinematically
 assigned to the thin (crosses) and thick disk (filled circles). The filled 
 circles surrounded with an additional circumference are the so-called 
 {\it TKTA} stars; they show thin-disk abundances but thick-disk kinematics. 
 Adopted from
 Reddy et al. (2006).}
   \label{alfe}
\end{center}
\end{figure}

Looking closely at the chemical compositions, a sharp distinction
between the two disks has become evident in the abundances of many elements, 
such as the $\alpha$-capture nuclei (O, Mg, Si, S, Ca, Ti). This is illustrated
in Fig. \ref{alfe}, borrowed from Reddy et al. (2006).
Some argue there exists genuine transition objects, but not all 
studies find them in their samples.

Galactic disks can become thinner by dissipative collapse while 
they are still rich in gas. Such straightforward connection between
the thick and thin disks does not seem viable in the light of 
the distinct chemical patterns that separate the two Milky Way disks.
Stellar disks can also become thicker with time, due to internal
(scattering and other dynamical interactions in the disk)
or external (satellite accretion) mechanisms. But again, such simple
path does not match the distribution of ages.

Two scenarios recently proposed in the literature appear 
feasible. Modelers have  argued for some time that mergers could have
been responsible to produce the thick disk, perhaps disrupting 
a previously existing disk, but allowing the thin disk to form 
and evolve independently afterwards. While the accretion of dry 
(gasless) system(s) may lead to features that clash with existing
observations (think rings of stars and thick-disk
characteristics that vary with galactocentric distance), the
acquisition of gas rich systems and on-the-fly formation of stars 
seems to work well (Brook et al. 2005, 2007).
The second scenario is based on new, relatively simple, models which indicate
that the observed distributions of metallicity, age, and $\alpha$/Fe
ratios could emerge as a result of the natural density gradient and 
the associated variation of the star formation rate with 
galactocentric distance, when coupled to radial mixing of  
stellar orbits (Sch\"onrich \& Binney 2008, 2009).

\section{The disk beyond the solar neighborhood}

The work in the solar neighborhood can now be complemented with observations
of more distant stars and stellar systems. These large-range data sets
will be most useful to discriminate among the proposed formation
scenarios for the thick disk and its connection to thin disk and 
halo.

Data on individual stars from the SDSS, which now accumulates close to
0.5 million stellar spectra, can tell us about abundance and kinematics
of stars over a wide range of distances, from tens to hundreds of
pc using low-mass stars, to tens of kpc for bright giants. An interesting
hint from SDSS is that the median of the metallicity distribution of 
the thick disk, about [Fe/H]$= -0.7$ dex, does not seem to vary between 4 $<R<$ 14 kpc, 
in distinct contrast with observations for thin disk stars, where significant gradients
are found for multiple elements using different tracers
(see Fig. 13 in Allende Prieto et al. 2006).

Both Cepheids (Andrievsky et al. 2002, 2004) and giants in 
open clusters (Yong et al. 2006) allow 
tracing the abundances of many elements at large galactocentric distances. 
A remarkable outcome of these studies is 
that the well-known thin disk abundance gradient (see the review by Maciel in
these proceedings) may flatten out
at $R> 12$ kpc (but see Sale et al. 2009 a for discrepant voice). This has
been suggested to indicate a flat density profile in the inner stellar halo
(Cescutti et al. 2007). Interestingly
enough, these studies also show that the $\alpha/Fe$ ratios increase
with galactocentric distance -- and so do the ratios of lanthanum (an $s-$process 
tracer) as well as europium (an $r-$process tracer) to iron.

The disk is by no means flat, and a better understanding of its 
structure is needed, in particular the flare and warp traced  
by stars (L\'opez-Corredoira et al. 2002; Momany et al. 2006)
 gas (Kalberla et al. 2007; Levine et al. 2006),
and dust (Drimmel \& Spergel 2001).
Infrared spectroscopic observations of vast numbers of 
red giants across the disk should provide much insight. 
APOGEE, part of SDSS-III, plans to obtain high-resolution $H$-band
spectra for 10$^5$ stars with a signal-to-noise ratio approaching 100 
between 2011 and 2014 (Allende Prieto et al. 2008). 
Preliminary studies suggest that more than 15
chemical elements can be sampled within the $H$ band, where  
dust obscuration is 5 times less than in $V$.

\section{Closing remarks}

The thin disk of the Galaxy likely fits in the overall bottom-up galaxy 
formation scenario in a $\Lambda$CDM universe, but a detailed picture of
its formation is still missing.
The stellar population of the thin disk is rich in kinematic structure, 
but appears chemically well-mixed.
The two statements in the previous sentence need not be in contradiction, 
as fine structure is likely missed due to limited abundance 
precision (currently $\sim$0.05 dex), and especially if most of the structure 
has a dynamical origin, excited by resonances and/or (modest) accretion.

The solar neighborhood needs to be placed in the context 
of the whole Galactic disk. Massive surveys of faint stars will do that,
and they will happen over the next 5--10 years.

Among the most pressing questions in this field, we could single out:
What is has been the star formation history of the 
solar neighborhood? (Must consider radial mixing!).
Which process(es) are mainly responsible for the 
stellar ÔclusteringÕ in phase space and the disk 
heating? What is the connection between the thin and thick disks?

The tools to address these questions are already in place:
 global astrometry methods, accurate spectroscopy from efficient instruments, 
detailed chemical analysis techniques (ÔisothermalÕ samples, larger 
samples, refined analyses), chemo-dynamical modeling, improved 
statistical techniques, and last but not least, data, to be 
provided by SDSS, RAVE, Gaia, and other supporting facilities.

\acknowledgements{I am grateful to the IAU, a Sociedade Astron\^omica 
Brasileira, and UCL graduate school for their generous support. Thanks go 
also to Katia Cunha for her warm hospitality,  
to Vivien Reuter for answering inquiries swiftly and 
with a smile, and to Mart\'{\i}n L\'opez Corredoira 
for stimulating discussions and comments on a draft.}

\end{document}